 \documentclass[onecollarge,natbib]{svjour2}

 \usepackage{graphicx}
\usepackage{epsfig}
\usepackage{natbib} 

\def\nl{\noindent}

\newcommand{\be}{\begin{equation}}
\newcommand{\ee}{\end{equation}}

\newcommand{\bee}{\begin{eqnarray}}
\newcommand{\eee}{\end{eqnarray}}

\newcommand{\feynslash}[1]{\slash\!\!\! #1}  

\bibpunct{[}{]}{;}{n}{}{,} 
\smartqed  
\usepackage{graphicx,amsmath,amssymb,bm}
\usepackage{color}
\usepackage{enumerate,array}

\usepackage{multicol}

\definecolor{navyblue}{rgb}{0.3,0.3,1}
\definecolor{purple}{rgb}{0.6,0,0.5}
\usepackage[colorlinks=true, pdfstartview=FitV, 
linkcolor=purple, citecolor= navyblue, urlcolor=navyblue]{hyperref}

\journalname{Few Body Systems}

\begin{document}

\title{Insights into the quark-gluon vertex from lattice QCD and meson spectroscopy    
\footnote{Presented by E. Rojas at LIGHT-CONE 2014,~May 26--30,~NCSU-USA}}  
\author{E. Rojas$^{1,2}$, B.~El-Bennich$^{1,3}$, J.~P.~B.~C.~de~Melo$^1$, and M.~Ali.~Paracha$^{1,4}$}

\institute{ 
$^1$Laboratorio de F\'isica  Te\'orica e Computacional, Universidade Cruzeiro do Sul, S\~ao Paulo, Brazil.\\
$^2$ Instituto de F\'isica, Universidad de Antioquia, Calle 70 No. 52-21, Medell\'in, Colombia.\\
$^3$ Instituto de F\'isica Te\'orica, Universidade Estadual Paulista,  S\~ao Paulo, SP, Brazil.\\
$^4$ Department of Physics, School of Natural Sciences, National University of Science and Technology, \\
  \hspace*{1,3mm} Islamabad, Pakistan.\\
}

\date{Version of \today}

\maketitle

\begin{abstract}
By comparing  successful quark-gluon vertex  interaction models  with the corresponding interaction extracted  from lattice-QCD data
on the quark's propagator,  we identify  common qualitative features which could be important to tune future interaction models beyond 
the rainbow ladder approximation. Clearly, a quantitative comparison is conceptually not simple, but qualitatively  the  results 
suggest that a realistic interaction should be relatively broad with a strong support at about $0.4-0.6$~GeV and infrared-finite.

\end{abstract}

\section{Introduction}
\label{intro}
The study  of the nonperturbative quark-gluon vertex has been of great interest over the past decade~\cite{Davydychev:2000rt,Skullerud:2002ge,
Skullerud:2003qu,Bhagwat:2004hn,Bhagwat:2004kj,Skullerud:2004gp,Matevosyan:2006bk,Kizilersu:2006et,Alkofer:2008tt,Hopfer:2013np,
Rojas:2013tza,Williams:2014iea,Gomez-Rocha:2014vsa,He:2013jaa,Binosi:2014aea} and  even though its tensor structure is well understood, a long way remains
to determine the corresponding form factors. Complete solutions of the Dyson-Schwinger equations (DSE) for the quark-gluon vertex are so
far out of reach, though tractable calculations can be realized by introducing model-dependent form factors~\cite{Roberts:1994dr,Alkofer:2000wg,
Maris:2003vk,Fischer:2006ub,Bashir:2012fs,Eichmann:2014xya}. One exciting feature of the DSE is that they provide a bridge between 
phenomenological models and the basic objects of quantum field theories, namely the Green functions which can be calculated with lattice-QCD 
simulations. 

In the Rainbow-Ladder~(RL) truncation of the DSE, in which the vertex structure is dictated by the perturbative limit, many hadronic properties 
of light mesons ($\lesssim 1$~GeV), quarkonia and the nucleon have successfully been described~\cite{Bashir:2012fs}. This suggests that to some 
extent the simple RL vertex structure is sufficient to calculate masses and decay constants of flavorless mesons when a judicious interaction model 
or dressing function is employed.   Direct extraction of the dressed quark-gluon vertex functions from lattice
simulations is not straight forward, as form factors associated with the longitudinal and transverse components have  been obtained 
in lattice simulations for different kinematic configurations, yet the range of space-like momenta on the lattice is rather limited and 
their use in numerical DSE applications is currently  impracticable. However, in computing the quark two-point function, lattice simulations 
do include, at least partially in quenched calculations, the effect of nonperturbative quark-gluon dressing. We made use of this in 
Ref.~\cite{Rojas:2013tza}, where we extracted an effective quark-gluon vertex by numerical inversion of the quark DSE using lattice 
data for its dressed mass function; see Eq.~\ref{ghostvertex} below.

In the present study, we also make use of an effective coupling strength,  $g_\mathrm{eff}^2$, which under certain circumstances  
allows  to extract  important qualitative information from  lattice data, and compare it with the corresponding effective interactions 
of phenomenological models in RL approximation~\cite{Munczek:1983dx,Frank:1995uk,Maris:1997hd,Maris:1997tm,Maris:1999nt}.
The DSE for the quark propagator and arbitrary flavor is given by,
\begin{equation}
   S^{-1}(p)  =   \, Z_2 (i\, \feynslash  p + m^{\mathrm{bm}}) +
   Z_1\, g^2\!\! \int_k \!  D^{\mu\nu} (q) \frac{\lambda^a}{2} \gamma_\mu S(k)\frac{\lambda^a}{2}\Gamma_\nu (k,p) \  ,
\label{DSEquark}
\end{equation}
where $q=p-k$, $\int_k\equiv \int_0^\Lambda\frac{d^4k}{(2\pi)^4}$ denotes a Poincar\'e invariant regularization and $Z_1$ and $Z_2$
are the vertex and quark wave-function renormalization constants.  In Landau gauge, the gluon propagator is transverse,
\begin{equation}
   D_{\mu\nu}(q)=T_{\mu\nu} (q)\Delta(q^2) \ ,
\end{equation}
with the projection operator $ T_{\mu\nu} (q) :=  g_{\mu\nu} -  q_\mu q_\nu/q^2 $. In QCD, the quark-gluon vertex  satisfies 
the Slavnov-Taylor-Ward Identity~(STWI) which can be well approximated by a ``ghost-improved'' Ball-Chiu 
vertex~\cite{Rojas:2013tza,Aguilar:2010cn,Aguilar:2013ac,El-Bennich:2013yna} :
\begin{equation}
\label{ghostvertex}
  \tilde \Gamma_\mu = \tilde X_0 (q^2)\,  F(q^2)\, \Gamma^\mathrm{BC}_\mu(k, p), 
\end{equation}
where $\tilde X_0(q^2)$ and $F(q^2)$ are vertex and ghost  dressing functions~\cite{Rojas:2013tza}, respectively, and 
$\Gamma^\mathrm{BC}_\mu$  is the Ball-Chiu (BC) vertex~\cite{Ball:1980ay}.  In the RL approximation and Landau gauge, the gluon  and  
vertex dressing functions  can be  absorbed   into the  definition, 
\begin{equation}
  Z_1 g^2 \, D_{\mu\nu} (q) \,  \Gamma_\mu (k, p) \ \longrightarrow \  \frac{\mathcal{G} (q^2)}{q^2} \, T_{\mu\nu} (q) \,  \gamma_\mu \ ,
  \label{vertexapprox}
\end{equation}
and the effective coupling is the sum of two terms which describe the nonperturbative and perturbative regimes of the interaction~\cite{Qin:2011dd}:
\begin{equation}
\label{qinchang}
  \frac{ \mathcal{G} (q^2)}{q^2} =   \frac{8\pi^2}{\omega^4}  D  \exp \left ( -  \frac{q^2}{\omega^2} \right ) +
             \frac{8\pi \gamma_m\, \mathcal{F}(q^2) }{\ln \left [\tau + \left (1 + q^2/\Lambda_\mathrm{QCD}^2 \right )^2 \right ] } \ .
\end{equation}
In Eq.~(\ref{qinchang}), $\gamma_m = 12/(33 - 2N_f )$, $N_f = 4$, $\Lambda_\mathrm{QCD} = 0.234$~GeV;
$\tau = e^2 - 1$; and $\mathcal{F}(q^2) = [1 - \exp(-q^2/4m^2_t  ) ]/q^2$, $m_t = 0.5$~GeV. 

The combined framework of DSE and Bethe-Salpeter equations (BSE) has been successfully used
to describe the pseudoscalar and vector meson spectrum, related weak decay constants and electromagnetic
form factors~\cite{Bashir:2012fs}. For flavored observables with light-quark current masses, $q=u, d, s$, this continuum QCD 
approach successfully reproduces the semi-leptonic decays $K^{+}\to \pi^{0}\ell^{+}\nu_{\ell}$ and 
$K^{0}\to \pi^{-}\ell^{+}\nu_{\ell}$ which proceed via a flavor changing 
charged current with a propagating spectator quark in the impulse approximation~\cite{Ji:2001pj,Maris:2002mz,Ivanov:1997yg}. The same is 
not true for heavy-to-light transition form factors and in general for heavy-light meson observables 
where one relies more on modeling~\cite{ElBennich:2009vx,ElBennich:2012tp,Ivanov:2007cw}. 
In such cases, an improvement of the theoretical framework is
direly needed~\cite{Ivanov:2000bm,ElBennich:2008qa,Souchlas:2010zz,Fischer:2009jm}.
The matrix element $\langle\pi^{0}(p)|\bar s\gamma_{\mu}u|K^{+}(k)\rangle$ can be used to determine physical observables such as 
branching fractions and the weak CKM matrix element $V_{us}$. The transition amplitude for the said decay can be completely described by two 
Lorentz vectors ($  t=Q^2$),  
 \begin{eqnarray}
  \langle\pi^{0}(p)|\bar s\gamma_{\mu}u|K^{+}(k)\rangle = f_{+}(t)\, K_{\mu}+f_{-}(t)\, Q_{\mu}\label{1} \ ,   
 \end{eqnarray}
where  $K_{\mu}=(p+k)_{\mu}$ and $Q_{\mu}=(k-p)_{\mu}$ are, respectively, the sum and difference of the momenta of the initial and final-state 
mesons, $t= -Q^2$  and  $f_{\pm}(t)$ are scalar transition form factors. 

Part of the DSE-BSE program is concerned with the calculation of hadronic observables  and meson spectroscopy by means of a single quark-gluon 
interaction vertex.  In order to accomplish this program, it is necessary and important  to identify phenomenologically successful vertex models 
beyond the RL approximation. Our primary interest in this work is to provide some qualitative criteria in the construction of these models.

\section{Effective coupling strength}

\begin{figure}[t!]
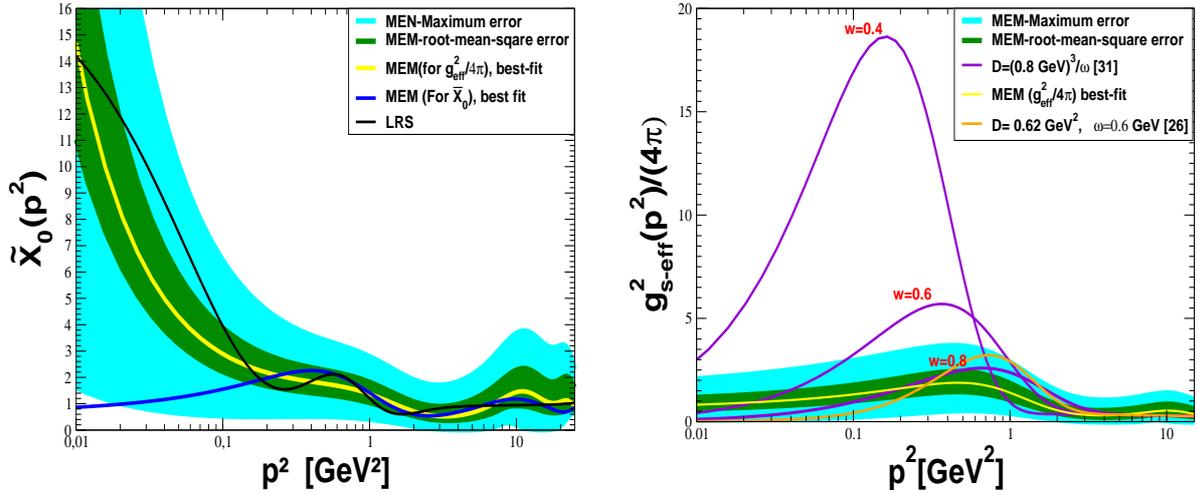

\centerline{
\epsfig{figure=x0-plot-mem-new.eps,width=7.50cm,height=6.50cm} 
\hspace{0.4444000cm} 
\epsfig{figure=to-light-cone.eps,width=7.50cm,height=6.50cm} 
}
\caption{
\textit{Left panel:} the effective quark-gluon dressing form factor, $\tilde X_0$, computed using the maximum entropy method~(MEM) and 
linear regularization (LRS); see Ref.~\cite{Rojas:2013tza} for details. The blue and yellow curves correspond to maximizing either 
the entropy of $\tilde X_0(p^2)$ or $g_\mathrm{eff}^2(p^2)/4\pi$ in the MEM, where $g_\mathrm{eff}^2(p^2)$ is the effective coupling 
strength of the interaction~\cite{El-Bennich:2013yna}. The latter is infrared enhanced, as in this case $p^2\tilde X_0(p^2)$ is finite in the 
limit $p^2\rightarrow 0$.
\textit{Right panel:} the purple curves correspond three different parametrizations of the effective coupling strength $g_\mathrm{eff}^2(p^2)=\mathcal{G}$ 
with the interaction model in Ref.~\cite{Qin:2011dd} ($\omega D = (0.8~\text{GeV})^3$ in all cases); the orange curve corresponds to the effective 
coupling strength in the Maris-Tandy model~\cite{Maris:1999nt}; the yellow curve corresponds to our effective coupling,   
$g_\mathrm{eff}^2(p^2)=4\pi\alpha_s(\mu)\Delta(p^2) F(p^2)\tilde X_0 (p^2)$, with $\alpha_s (\mu)= 0.295$~\cite{El-Bennich:2013yna}.}
\label{fig1}
\end{figure}

It was shown in Ref.~\cite{Rojas:2013tza} that it is possible, within error bars,  to reproduce lattice-QCD data~\cite{Parappilly:2005ei} on the 
quark propagator with the ghost-improved BC vertex of Eq.~\eqref{ghostvertex}. It is also well known that in solving the DSE numerically, 
the BC vertex can be approximated by $\tfrac{1}{2} [A(k)+A(p)] \gamma_\mu$ such that the solutions for $A(p^2)$ and $B(p^2)$ 
are similar to those employing the full BC vertex. This prompts us to compare our ghost-improved vertex model with common RL models 
focusing on effective coefficient of the vertex structure $\gamma_\mu$ in the DSE kernel. In this approximation, we can define the 
effective coupling strength $g_\mathrm{eff}^2(q,p)$ in either case, RL or BC, as in Eq.~\eqref{vertexapprox},
\begin{equation}
  Z_1 g^2 \, D_{\mu\nu} (q) \,  \Gamma_\mu (k, p)  \ \longrightarrow   \  \frac{g_\mathrm{eff}^2 (q^2)}{q^2} \,  T_{\mu\nu}\,  \gamma_\mu \ ,
\end{equation}
which corresponds to 
\begin{equation}
\label{eq:geff}
\frac{g_\mathrm{eff}^2 (q,p)}{q^2}=
  \begin{cases}
   &\mathcal{G} (q^2) \quad   \text{[RL]} \, \\    
   \\
   & 4\pi\alpha_s(\mu)\Delta(q^2) F(q^2)\tilde X_0(A(k)+A(p))/2   \quad  \text{[BC]}  \\       
 \approx 
   & 4\pi\alpha_s(\mu)\Delta(q^2) F(q^2)\tilde X_0  \ .
   \end{cases}
\end{equation}
In Eq.~\eqref{eq:geff}, we made use of the heavy-flavor approximation, $\tfrac{1}{2} [ A(k)+A(p) ] \approx 1$, in the BC vertex, so that 
$g_\mathrm{eff}^2 (q^2)$ only depends on $q^2$. It is not worth introducing this factor as part of the effective coupling strength,
since it is flavor dependent and this approximation can be seen to be reasonable from Fig.~3 of Ref.~\cite{Rojas:2013tza}. In there, the
variation, $1\lesssim (A(k)+A(p))\frac{1}{2} \lesssim 1.5$, in the range, $(0.14~\text{GeV})^2 \lesssim  p^2  \lesssim  (4.3~\text{GeV})^2$,
is for the light flavors whereas for heavy quarks this factor is close to one. The difference in the effective coupling strength between 
$\omega = 0.4$~GeV and $\omega = 0.8$~GeV is huge; thus, roughly speaking our conclusions are independent of these less significant details. 

\begin{table}[t!]
\caption{Mass spectrum and decay constants for flavor-singlet and nonsinglet $J^{P}=0^-$ mesons, where we follow 
Particle Data Group conventions~\cite{Beringer:1900zz}. Both models refer to the interaction ansatz in Ref.~\cite{Qin:2011dd}, 
where we use the values $\omega = 0.4~\text{GeV}$ and  $\omega D =(0.8~\text{GeV})^3$  for Model~1 and $\omega =0.6~\text{GeV}$ 
and $\omega D =(1.1~\text{GeV})^3$ for  Model~2. Dimensioned quantities are reported in GeV and reference values 
in the last column include experimental averages. }
\label{table1}
\begin{center}
\begin{tabular}{p{1.5cm}p{2cm} p{2cm}p{2cm}}
\hline\hline
                      & Model 1      &   Model 2       &     Reference     \\
\hline
$m_{\pi}$             &  0.138          &     0.153               &  0.139~\cite{Beringer:1900zz}             \\
$f_{\pi}$             &  0.139          &     0.189               &  0.1304~\cite{Beringer:1900zz}            \\ \hline
$m_{\pi(1300)}$       & 0.990                & 1.414              &  $1.30\pm 0.10$~\cite{Beringer:1900zz}    \\
$f_{\pi (1300)}$      & $-1.1\times10^{-3}$  & $-8.3\times10^{-4}$&                                           \\
$m_K$                 & 0.493                & 0.541              &  0.493~\cite{Beringer:1900zz}             \\
$f_K$                 & 0.164                & 0.214              &  0.156~\cite{Beringer:1900zz}             \\
$m_{K(1460)}$         & 1.158                &1.580               &   1.460~\cite{Beringer:1900zz}             \\
$f_{K(1460)}$         & $-0.018$             & $-0.017$           &                  \\
$m_{\eta_c(1S)}$      & 3.065                & 3.210              &  2.984~\cite{Beringer:1900zz}              \\
$m_{\eta_c(2S)}$      & 3.402                & 3.784              &  3.639~\cite{Beringer:1900zz}              \\
\hline \hline
\end{tabular}
\end{center}
\end{table}

Despite the qualitative nature of our analysis, we may derive some insight on the necessary
domain of support of the interaction, i.e. the combined effect of gluon- and vertex-dressing functions, from comparison with the
meson spectroscopy produced by the interaction models and the interaction we extracted from the lattice simulation data, i.e.
both ans\"atze in Eq.~\eqref{eq:geff}. From Table~\ref{table1} it is clear that for $\omega = 0.6$~GeV and $D\omega = (1.1~\text{GeV})^3$
the mass values of the exited states $\pi(1300)$, $K(1460)$ and $\eta_c(2S)$ are in better agreement with the average experimental
data than for $\omega = 0.4$~GeV. However, the interaction with $D\omega = (1.1~\text{GeV})^3$ and $\omega= 0.6$~GeV does not 
yield acceptable decay constants and overestimates the masses of the ground states. A better interaction candidate may be found 
in the choice $\omega = 0.6$~GeV and $D= (0.8~\text{GeV})^3/\omega$, yet we encounter numerical complications for these 
parameter values  when we solve the DSE in 
the complex plane\footnote{The quark propagators exhibits pairs of complex conjugate singularities which complicate the calculation.}.
For the time being this does impede the computation of  the 
masses of radial resonances.  Nonetheless, this problem can be overcome by using a Nakanishi representation of the propagators. 
The choice $\omega = 0.6$~GeV is also motivated by the fact that the Maris-Tandy model~\cite{Maris:1999nt} adequately reproduces 
experimental bottonium spectroscopy with the interaction parameters tuned to $D= 0.62$~GeV$^2$ and $\omega\sim 0.6$~GeV~\cite{Blank:2011ha}. 
This is particularly significant since for heavy flavorless mesons the RL approximation is appropriate and flavor effects in the 
effective coupling strength should be minimal in this case. 

In Fig.~\ref{fig1}, we depict the effective coupling strength of Ref.~\cite{Qin:2011dd} for three values of $\omega$ and 
$D= (0.8~\text{GeV})^3/\omega$ fixed. In there, some common qualitative features may be appreciated; note in particular that the 
effective interaction is finite in the infrared. On the other hand, meson spectroscopy does  not  allow to draw general conclusions 
about the finiteness of the interaction. Modern DSE and lattice-QCD studies find that the gluon propagator is a bounded, regular function 
of spacelike momenta, which achieves its maximum value on this domain at $q^2=0$~\cite{Binosi:2014aea}. Nonetheless, it is necessary 
to establish whether the vertex possesses some structure which can qualitatively alter this behavior as the effective 
interaction depends on the product of gluon propagator and vertex dressing functions.
In general, it is quite difficult to put absolute constraints on  $D\omega$ in the range $\omega \sim (0.4-0.8)$~GeV. 
However, once $D\omega$ is fixed the best values for the masses of the radial resonances correspond to 
a broader distribution in momentum space or equivalently larger values of $\omega$~\cite{Popovici:2014pha}. Taking also into account 
that a broader distribution with a less pronounced peak is also required to reproduce the heavy-quarkonia spectrum, this suggests 
that a suitable interaction for a wide  range of applications beyond the RL ought to have a strong and broad support concentrated 
at about $0.2-0.3$~GeV$^2$. 

A deeper insight is gained from the lattice data by inspection of the  functional behavior of $\tilde X_0(q^2)$. All the solutions for $\tilde X_0$ 
in Fig.~\ref{fig1} are consistent with the lattice data for $A(p^2)$ and $B(p^2)$ as can be seen in Fig.~13 of Ref.~\cite{Rojas:2013tza}. 
However, it also becomes clear from Fig.~14 in the same reference that {\em only\/} when one solves the quark DSE with an infrared enhanced 
form factor $\tilde X_0$, the values of $A(p^2)$ and $B(p^2)$ become consistent with lattice QCD data at low momenta. This suggests 
an inflection of $\tilde X_0$ towards a finite limit for $q^2 \to 0$~GeV$^2$. Therefore, even though lattice data for the dressed-quark wave 
and mass functions do not strongly enough constrain the infrared behavior of $\tilde X_0(q^2)$  in the DSE inversion process, we find 
that requiring consistency between the DSE solutions and lattice QCD results provides a valuable source of information and constraints 
on the infrared domain.

\section{Conclusions}

By comparing experimental data on the pseudoscalar meson mass spectrum with realistic model calculations in the RL approximation,  
one deduces that the data prefer a broader distribution in momentum space for the effective coupling strength~\cite{Blank:2011qk,Rojas:2014aka,
Popovici:2014pha}, $g_\mathrm{eff}^2$, rather than a localized and sharply peaked interaction. An effective interaction extracted 
from lattice QCD data on the quark propagator is consonant with such a distribution and also points at an infrared-finite limit of 
$g_\mathrm{eff}^2$. Self-consistency requires thus an infrared-finite functional behavior of $\tilde X_0$. Nonetheless, in order to 
go a step further in finding an interaction ansatz valid for a wide range of mesonic observables, one may need to  resort to a 
more realistic quark gluon-vertex interaction and take into account the important structure of its transverse components. 
In the past two decades, some progress in this direction has been made~\cite{Bender:1996bb,Eichmann:2008ae,Fischer:2008wy,
Fischer:2009jm,Qin:2013mta,Heupel:2014ina,Rojas:2013tza} and phenomenological models should be extended to beyond the RL 
approximation using the guidelines of these studies as well as other input, such as the one proposed in the present study. 

\vspace{0.2cm}

\nl {\bf Acknowledgments.} This work was supported in part by the Brazilian agencies FAPESP (Funda\c{c}\~ao de Amparo \`a Pesquisa do
Estado de S\~ao Paulo) and CNPq (Conselho Nacional de Desenvolvimento Cient\'\i fico e Tecnol\'ogico).
We  benefited from access to the computing facility of the  Centro Nacional de Supercomputa\c{c}\~ao at the Federal University of Rio Grande do Sul (UFRGS).
 We thank  T.~Frederico and O.~Oliveira 
for their valuable collaboration. E.~R.  thanks 
financial support provided by ``Sostenibilidad-UDEA 2014–2015''  
and the organizer of the Light-Cone 2014 Conference for the invitation.


 \end{document}